# Enhanced Performance Neutron Scattering Spectroscopy by Use of Correlation Techniques

F. Mezei[1,2,3*], M. T. Caccamo[3], F. Migliardo[4,5], S. Magazù[3,6]

[1]*European Spallation Source ERIC, Lund, Sweden*
[2]*HAS Wigner Research Center, Budapest, Hungary*
[3]*Department of Mathematical and Informatics Sciences, Physical Sciences and Earth Sciences, University of Messina, Italy*
[4]*Department of Chemical, Biological, Pharmaceutical and Environmental Sciences, University of Messina, Italy*
[5]*Institute for Integrative Biology of the Cell (I2BC), CEA-CNRS-Université Paris Sud, Orsay, France*
[6]*LE STUDIUM, Loire Valley Institute for Advanced Studies, Orléans & Tours, France and Centre de Biophysique Moléculaire (CBM,CNRS), Orléans, France*

\* ferenc.mezei@esss.se

**Abstract**

In neutron correlation spectroscopy pseudo-random modulation of the incoming beam intensity is used to determine the velocity distribution of the detected neutron spectrum. The intensity of the modulated beam exceeds the beam intensity in direct spectroscopy by up to two orders of magnitude. On the other hand, the propagation of the counting noise in the correlation algorithm of data reduction is disadvantageous for the lowest intensity parts of the scattering spectrum. We propose an innovative approach for mitigating this drawback in particular for vibrational spectroscopy at high intensity pulsed neutron sources such as the European Spallation Source (ESS). The new design is based on two dimensional time-of-flight data collection combining low resolution direct spectroscopy on the basis of the source pulse structure and high resolution correlation spectroscopy with the help of a state-of-the-art statistical neutron beam chopper. We have identified a formulation of the data processing algorithm by matching the data processing channels to the inherent time resolution of the statistical chopper that makes the reconstruction of the direct time-of-flight spectra mathematically exact and independent of all other contributions to the instrumental resolution. In order to provide flexibility and versatility for the instrument novel methods of trading intensity vs. resolution without changing the statistical chopper or its speed are also proposed. These either imply changing simple beam window diaphragms or varying the intensity and resolution options in the data reduction algorithm in the data processing after the experiment. This latter innovative feature is a unique and very promising potential of the correlation method in contrast to conventional direct data collection. A further important advantage of correlation spectroscopy is the reduced sensitivity to external background, also due to the much higher recorded signal intensity. This is particularly advantageous for extending the operational range for epithermal neutrons arriving just after a source pulse. In the suggested configuration for hot and thermal neutron vibrational spectroscopy, with design parameters matched to the desired energy resolution envisaged by the conventional chopper approach, the correlation chopper variant allows us to achieve very significant gains in terms of spectrometer efficiency. In particular, the high intensity for the most intense features in the spectra and the reduced sensitivity to sample independent background make correlation spectroscopy a powerful choice for studying small samples.

**Keywords:** Inelastic Neutron Scattering, statistical chopper, correlation spectroscopy, vibrational spectrometer

# 1. Introduction

Neutron spectroscopy is mainly aimed at the characterization of the kinds and strength of intra- and inter-molecular interactions which determine both the configuration and the dynamics of atoms and molecules in condensed matter systems. For molecular systems the energy range to be explored is particularly large and challenging, and the domain of interest can reach from a few to several hundred meV. This for example represents an evolution of the approach of such studies which initially addressed just the understanding of the vibrational modes involving hydrogen atoms. Boosting the instrumentation improvement and the upgrading of calculation and modelling performances are key components of progress.

The new chances offered by the last generation neutron sources such as the European Spallation Source (ESS) open up qualitatively new opportunities by combining order of magnitude higher beam intensities with enhanced innovative approaches in designing flexible and versatile neutron spectrometers whose resolution could be competitive with those of the vibrational optical techniques and whose performances should allow us to span a wide range of experimental conditions as well as of scientific and technological fields.

Inverted geometry time-of-flight (TOF) spectrometers form a successful class of instruments at current pulsed neutron sources based in Europe, US and Japan. These rely on the precise determination of the incoming neutron velocity/energy with the help of the time elapsed between the source pulse and the detection, and the analysis of the final neutron energy is performed by crystal analyzers before the detectors. The TOSCA, IRIS, OSIRIS spectrometers at the ISIS facility (Chilton, UK) and BASIS and VISION spectrometers at the SNS, Oak Ridge National Laboratory (Oak Ridge, US) are very successful examples. These instruments use beams from so called de-coupled moderators at short pulse sources. When using higher intensity coupled moderators with about 0.5 ms response time for slow neutron emission at short pulse sources or at long pulse source in general, pulse shaping choppers are needed to achieve or surpass the resolution offered by the shorter pulses emitted by the de-coupled moderators. An already successfully example is the spectrometer DNA at J-PARC and pulse shaping by neutron choppers in order to assure required resolution is common for many instruments planned at ESS. Our current study focuses more closely on the use of statistical choppers for pulse shaping for high and variable resolution inverted geometry spectrometers, which often also incorporate diffractometry. Our considerations and results, including a re-formulation of data processing are also relevant for the broader perspectives of the use of correlation spectroscopy in neutron scattering, and offer significantly enhanced performance for vibrational spectroscopy at a long pulse source such as ESS.

The correlation technique [1-7], which was developed starting from the late sixties in connection with stationary sources and recently adapted to the new generation of pulsed spallation neutron sources, offers the great advantage to optimize the use of the available neutron flux through modulation in time in a pseudo-random way. In such a case the system scattering function can be reconstructed by performing the cross-correlation of the measured signal with the modulation sequence. Thanks to such an approach, in comparison with traditional TOF spectrometers, where 1-5% of the incident time average beam intensity can be employed, one can use up to 50%. As we will see below, the noise propagation in correlation techniques is more favorable at pulsed sources than at continuous ones and one can anticipate further high performance developments at pulsed sources.

# 2. Statistical chopper and correlation spectroscopy

In Figure 1 a simplified diagram of a correlation technique based instrument is shown; here, the neutron beam is time-modulated by a statistical chopper following the pattern $M(t)$ with $0 \leq M(t) \leq 1$, while the intensity scattered by the sample is registered at a detector.

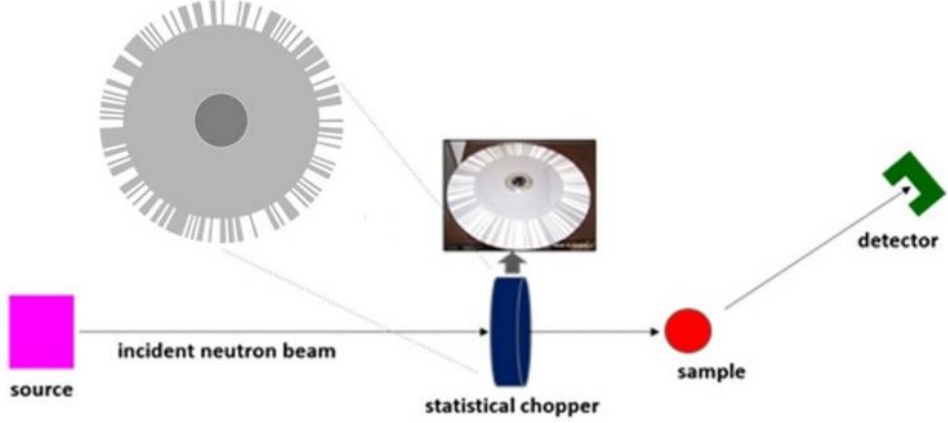

**Figure 1:** Simplified correlation technique based instrument

To obtain a statistical chopper one has to implement a pseudo-random sequence $n_i = 0$ or $1$ which has the property

$$\sum_i n_i n_{i-j} = c\delta_{0,j} + d \tag{1}$$

where $c$ and $d$ are constants. Note that the sequence $n_i$ is to be understood as repeating itself periodically over all integers $i$, and accordingly many relations need to be considered as $mod(n)$, where $n$ is the number of elements in the sequence. It is practical to define $n_i^*$ as $n_i^* = 1$ if $n_i = 1$ and $n_i^* = -1$ if $n_i = 0$. With this definition we can make the auto-correlation function $F_j$ a $\delta$ function:

$$F_j = \sum_i n_i n_{i-j}^* = c\delta_{0,j} \tag{2}$$

The example reported in the table on the right of Figure 2 is constituted by $n = 255$ elements $i$, which is a number of the form $2^n - 1$ where the integer $m$ is equal to 8 and $c = 127$. The corresponding statistical chopper design is reported in Figure 2 (on the left). This sequence has been chosen for the new CORELLI spectrometer at SNS, Oak Ridge [7]. The number of practical choices of random sequences for neutron scattering applications are limited [3,8]. For about equal number 1 and 0 elements we also have $n = 127$ and $n = 511$ available. Precision manufacturing for the 255 unit has been accomplished for CORELLI, which operates well at 18000 RPM rotation speed. The 511 sequence might be an important challenge for similar speed, although chopper discs with > 1000 slits have been accomplished for so called Fourier choppers [9].

If we consider the beam just after the chopper at any point $r$ of the beam window, the transmission modulation will depend on the position $r$ as the chopper slits move across the beam. This time dependence can be expressed by a position dependent delay $t(r)$. Neutrons crossing the chopper at point $r$ will be detected with a further time delay $t(T_r)$, where $T_r$ stands for all parameters describing a specific neutron trajectory passing by point $r$ (positions, path, velocity, scattering and detection events, etc.). The key point is that for any $r$ and $T_r$ the pattern $n_i$ in Figure 3 will contribute to the detector signal by an identical perfect binary modulation function with a delay $(t(r) + t(T_r))$. Thus if we measure the time in the chopper slit units $t_0$, we will recover exactly the pattern in Figure 3 with a time delay of $k = int((t(r) + t(T_r))/t_0)$ time channels. Thus the measured signal becomes:

$$M_i = \sum_k n_{i-k} \int_{k=const} P(t(r) + t(T_r)) dr dT_r = \sum_k n_{i-k} P_k \tag{3}$$

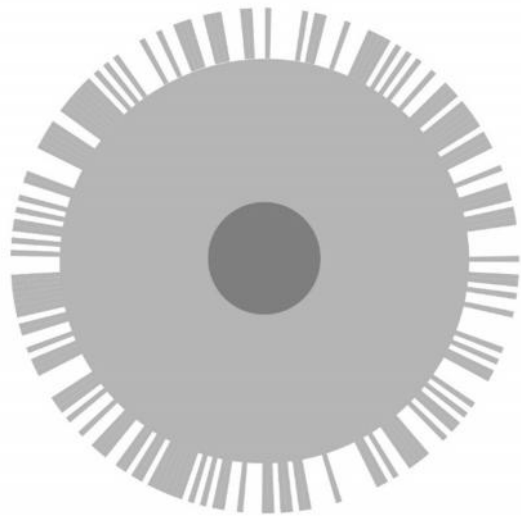

**Figure 2:** Statistical chopper (on the left) realized on the basis of the sequence reported in the table (on the right)

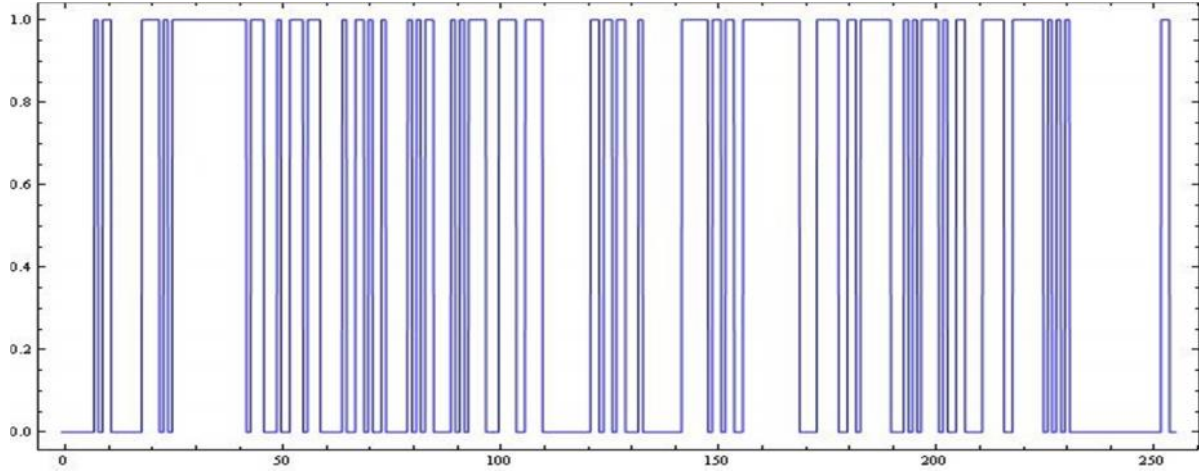

**Figure 3:** The pattern of the pseudo-random beam modulation function, which repeats itself periodically in time. Here the numbers on the x axis correspond to time channels of width equal to the unit slit width of the chopper. For CORELLI it is 1.412 ± 0.04° and the corresponding unit time at 18000 RPM rotational speed is $t_0 = 13.07$ μs. This time pattern applies identically to any point $r$ just after the chopper – anywhere in the beam cross section of any shape and size – with a time delay $t(r)$, depending on its position in view of the advancing of the chopper slits in the rotation and radial distance from the chopper axis.

where $P$ is the continuous distribution function of the time delays for different neutron paths with starting points at the statistical chopper $r$ and following trajectories $T_r$. Thus $M_i$ is the resulting discrete distribution function of neutron detection on the absolute time scale defined by the phase of statistical chopper operation in channels of length $t_0$. The final result of the measurement is the TOF distribution function $cP_j$ (where $c$ is the time average intensity gain vs. a single slit of unit width), which can be reconstructed in terms of the discrete units of the absolute time $t_i = it_0$ with integer $i$, by building the correlation function between the patterns $M_i$ and $n_i^*$ following eqn. (2):

$$F_j = \sum_i M_i\, n_{i-j}^* = cP_j \qquad (4)$$

…..Here $P_j$ is proportional to the sample scattering function $S(Q, \omega)$ modified by all the conventional instrumental response effects, in addition to the distribution of the chopper operation related time differences $t(\mathbf{r})$. The crucial point in this formulation is that it offers an exact correlation reconstruction algorithm with no approximation (within the geometrical precision of the chopper disc manufacturing), e.g. it does not depend at all on the actual neutron pulse shape formed by the interplay between chopper slit and beam window geometry. The reconstructed TOF spectrum $cP_j$ is exactly the multiple of the spectrum $P_j$ that would be observed if we would replace the statistical chopper with one with a single slit identical to the smallest one in the statistical chopper and leaving anything else unchanged. The beam window can be of any size and shape; it will only modify the global TOF distribution function $P_j$ in an exactly calculable manner, since the time difference $t(\mathbf{r})$ only depends on the chopper speed and the position $\mathbf{r}$ with respect to an arbitrary fix reference point $\mathbf{r}_0$.

The fact that the function $P_j$ can only be exactly determined for integer multiples of the statistical chopper unit time $t_0$ (the actual time resolution of the chopper) is another cornerstone of our formulation. It has no effect on revealing details of line broadening and line shift much smaller than this resolution: these will result in variation of the distribution $P_j$ between neighboring time channels $j$. In contrast to the common belief in TOF spectroscopy that the data collection time channel width must be considerable smaller than the instrumental TOF resolution in order to avoid unnecessary resolution broadening, this prescribed, mandatory time channel width $t_0$ in data acquisition and analysis leads to no loss of information or resolution here. Details on the line shapes on the scale of the resolution will be masked by the shape of the resolution function anyway for a model independent study. If we work with an assumed line shape, its parameters can as well be fitted /determined from a few points representing the integrals over well-defined elementary channels as from the same intensity divided over a large number of channels with correspondingly less accuracy for each.

An important practical consequence of the analysis via eqs. (2) and (3) is that increasing the size of the beam window alone, without any change of the chopper or its speed, will deliver the corresponding gain of beam intensity at the price of lower resolution. For the statistical chopper the time average beam transmission is independent of unit slit size and rotation speed. Reducing the resolution without changing the chopper speed only affects the dynamic range within one chopper rotation. This can be of no significance at a pulsed source, where the time measured from the source pulse allows for distinguishing between full rotations of the statistical chopper and tracking the $mod(n)$ effects.

### 3. Basics of correlation spectroscopy.

*3.1 Signal and Noise*

The basic idea behind all versions of beam modulation neutron spectroscopy (Neutron Spin Echo, Fourier choppers, statistical choppers) is to get around the Liouville theorem coupling between resolution and beam intensity, i.e. to achieve higher resolutions in the determination of the neutron velocity or velocity change than the monochromaticity of the neutron beam used. The intensity gains can be tremendous, e.g. up to 5 orders of magnitude in NSE. The price to pay is also common to these methods: the statistical error of the neutron data will correspond to the high integral beam intensity delivered, which can mask a small feature contained together with a strong one in the same scattered spectrum. This is of little concern in the most common uses of NSE for the study of the temporal decay of dominating features of correlations, where nearly all of the detected intensity comes from the studied phenomenon. In some cases the quasielastic structure of interest is combined with an elastic line and the statistical error of the elastic signal can make difficult to observe a weak inelastic signal on the top of it.

This basic phenomenon that historically led to the small share of correlation/modulation spectroscopy we can observe today in inelastic neutron spectroscopy, where the sustained successful applications after the great enthusiasm in the 1960's remained limited to the cases of modulation of the beam polarization [10]. In particular the by now discontinued IN7 correlation spectrometer at ILL revealed the difficulties due to the masking of the inelastic spectra by the counting statistics background of high intensity elastic contributions. This difficulty is absent if one is interested in the study of the dominating elastic or quasi-elastic part of the spectra, which is the case of CORELLI: to single out elastic diffuse scattering from inelastic contributions. Another emerging opportunity of this kind and of high interest is the case of the variable resolution elastic neutron scattering (RENS) spectroscopy [11-13], both on pulsed and continuous sources.

Pulsed neutron sources (in particular those with high intensity and therefore longer pulse structure) represent another area of high performance potential for modulation spectroscopy by the opportunity to separate neutron counts from parts of spectra and prevent the weak (e.g. inelastic) signals to mix in the data collection with the large (e.g. elastic) ones. One instructive and highly successful example is the Fourier diffractometer at the pulsed reactor in Dubna [9]. The about 300 μs pulses of this source are too long for high resolution TOF diffraction, but allow for separating the diffraction pattern into low resolution resolved groups. The superposed Fourier modulation method is then used to analyze each of these groups separately with a TOF resolution better than 10 μs under favorable conditions of counting statistics.

Indeed, in modulation techniques the common counting statistical error (in what follows referred to as "counting noise") for each channel of the final result, in the correlation/modulation reconstructed spectrum of interest, is given by the square root of the integrated intensity of the whole simultaneously collected modulated spectrum. If the spectrum consists of a dominating feature, e.g. an elastic line contributing to most of the total integrated intensity collected by the detector, the counting noise for this structure is essentially the same as in conventional spectroscopy: the square root of the detected intensity. For a lower intensity component in the spectrum, the counting noise will still correspond to the square root of the total detected intensity, reducing the data collection efficiency. In terms of counting/intensity figure of merit (FOM) this means an increase in counting time to catch up the counting noise increase. In what follows we will consider gains or FOM always in terms of required counting time to achieve a given signal to noise ratio. This time is inversely proportional to (signal)$^2$/noise.

…..A specific and simple example for statistical choppers is the counting noise for the contents of a time channel in reconstructed direct TOF spectrum. In view of eqn. (4) the counting noise of $cP_j$ will be given by the square root of the total counted intensity $\sum M_i$, since $|n_i^*| = 1$ for all $i$. Note that by definition $\sum M_i = c \sum P_j$, i.e. the error of the value of each channel of the reconstructed TOF spectrum is equal to the error of the contents of all channels added together. To turn this into a numerical example, if by the use of the modulation technique we gain a factor of 100 in beam intensity, and we are looking to the content of a reconstructed TOF spectrum channel $P_j$ with 1% of the total intensity of the detected spectrum, the counting FOM is 1, i.e. no gain, no loss in the case that the error only comes from counting noise of the signal. (We will discuss below a more beneficial feature of error propagation.) If there is in addition an external background, the intensity gain of the useful beam in the modulation method will reduce considerably or decisively the weight of the external background, as in some historic tour de force examples [14]. If the signal independent background is significantly larger than the signal we are looking for, even weak features will fully benefit of the beam intensity gain by the modulation method.

A crucial difference in the use of modulation methods on pulsed and CW sources is the opportunity that opens up to efficiently manage/shape the counting noise by a two dimensional TOF data recording. This implies that for every neutron counting event both the time with respect to the firing of the source $t_S$ and the time with respect to the position zero signal of the modulating device (e.g. chopper) are recorded [15]. For example on the Fourier diffractometer in Dubna, slices of the TOF diffraction pattern comparable to the length of the neutron pulse represent much less total intensity than the whole pattern. Thus the counting noise for every structure of the spectrum (Bragg peaks) in the Fourier analysis only comes from a few neighboring peaks, allowing for favorable relative weight within the slice of $t_S$ that fully (or nearly so) contains a given Bragg structure in the scattered beam.

*3.2 Time structures and assignments*

Figure 4 illustrates the basic time relations in correlation data collection at a pulsed source. It must rely on a two dimensional TOF data acquisition system, i.e. recording for every neutron detected both the time $t$ elapsed since the last start of the last chopper revolution and the time $t_S$ elapsed since the firing of the pulse [15]. In this variable space $(t, t_S)$ the statistical data collection time channels $t_i$ are defined as $i = int(t/t_0)$, with channel width $t_0$ as described above. The time channel width for the source pulse reference $t_S$ is rather arbitrary.

…..The "detector system" can be different depending of the type of experiment considered. In the case of a TOF diffractometer the detector system simply represents a detector at a given scattering angle and the sample is somewhere between the statistical chopper, actually without any effect to the diagram since the neutron velocity does not change in the scattering process. In the case of inverted geometry spectroscopy the detector system starts with the sample and includes the analyzer crystal and the actual detector at a given scattering angle. Indeed, since the neutron propagation time is a constant from sample to detector, one can only consider in the current analysis the neutron arrival times at the sample and the scatter of the propagation times in the crystal analyzer and detector are incorporated into the effective sample scattering distribution as a function of incoming neutron velocity. The generalization of the conclusions below is rather straightforward for other scattering configurations, such as direct geometry TOF spectroscopy as in CORELLI.

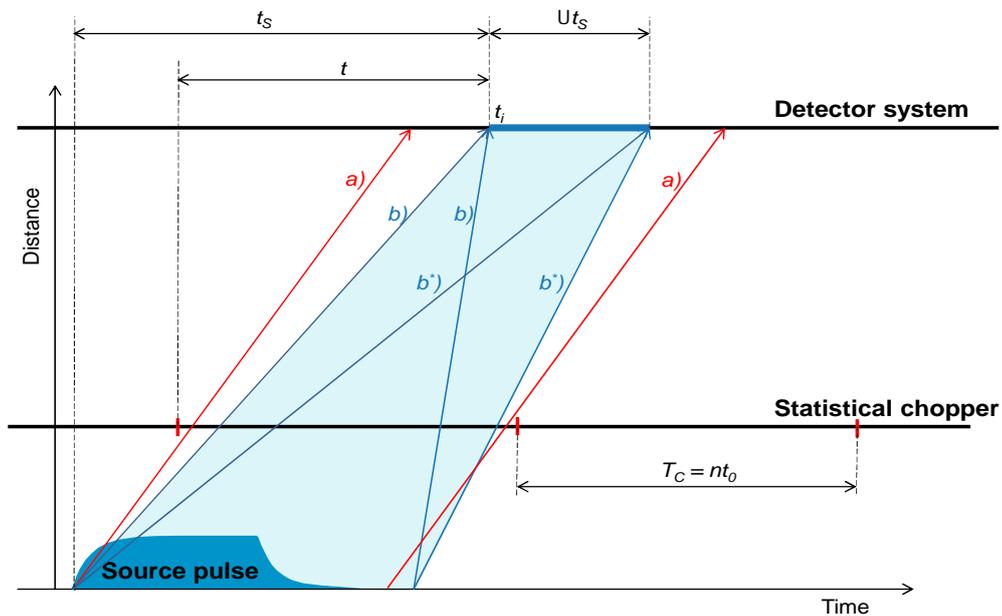

**Figure 4**. The elements determining data collection and correlation data reconstruction over a time domain of unlimited duration in correlation spectroscopy at a pulsed neutron source. The details are given in the text.

The statistical chopper and the source must run asynchronously, which can practically be very well achieved [16] by choosing a statistical chopper frequency close to a multiple of the source frequency within a small value $v \ll 1$ Hz. Thus within a data collection period of $1/\varepsilon$ the shape of the source pulse is sampled for integration in a good approximation using a large number of equal intervals for each statistical chopper time channel $t_i$, i.e. there remains no correlation between the contents of the correlation time channels $M_i$ and the shape of the source pulse. The so measured spectra $M(t, t_S) \equiv M_i(t_S)$ after a data collection period $> 1/\varepsilon$ will provide via eqn. (4) the contribution to the reconstructed TOF spectrum by neutrons detected at time $t_S$ from the source pulse. Lines *b)* in Figure 4) show the limits of neutron velocities that are contributing to the signal at a time $t_S$. It is important to observe that the trajectories within these limits cross the statistical chopper within a time interval smaller than the chopper rotation period $T_C$. If this condition is not met, spurious, (but fully predictable) "echo" signals can turn up in the reconstructed spectrum: for the statistical chopper data collection time channels are only defined $mod(n)$. Lines *a)* indicate the limits of the period in time $t_S$ of the contribution with significant intensity of a given velocity to the spectrum.

Before data evaluation/reconstruction along eqn. (4), one can freely add chopper time channels $t_S$ as long as all trajectories with not negligible intensity cross the statistical chopper within the duration of one period of rotation. This is shown in the figure for the case of adding all source pulse time channels between $t_S$ and $t_S + \Delta t_S$. As a consequence of the periodicity of the pseudo-statistical pulse sequence, the direct TOF spectrum reconstructed by the correlation calculations is only defined within an integer multiple of the chopper rotation period $T_C$. However, by knowing the smallest and/or largest flight times in the part of the spectrum covered by the combined collected data in the $t_S$ range selected (between trajectories *b)* and *b*$^*$*)* in the figure), one can determine the time domain covered between the smallest and largest time-of-flights from the chopper to the detector, corresponding to the two diagonal trajectories in the shaded area, respectively. Namely:

$$t_{TOF}^{max} = (t_S + \Delta t_S)\frac{d_{CD}}{d_{SD}} \qquad \text{and} \qquad t_{TOF}^{min} = (t_S - \Delta t_P)\frac{d_{CD}}{d_{SD}} \qquad (5)$$

where $d_{SD}$ and $d_{CD}$ are the distances of the detector from the neutron source moderator and the statistical chopper, respectively, and $\Delta t_P$ is the total neutron source pulse length including the tail down to negligible intensities. Thus knowing the $t_S$ domain is sufficient to identify the multiples of the chopper revolution period $fT_C$ ($f = 0, 1, 2, ...$), which need to be added to the time scales of the parts of reconstructed spectra in order to recover the reconstructed, direct TOF spectra on the absolute time scale for each, adequately selected $t_S$ domain. The absolute time scale here stands for the delay of the intensity modulation pattern at the detector with respect of the beam intensity modulation by statistical chopper. This process is very accurate due to the high precision of the statistical chopper rotational period. Finally, the so reconstructed direct TOF spectra over all the partial $t_S$ sub-domains constituting the whole time span between source pulses can be added without limitation, in order to improve statistics and/or to obtain the whole TOF spectrum. In the particular example in the figure the lower limit of TOF for the diagonal *b)* is $< T_C$, and the upper one – along the diagonal *b*$^*$*)* – is $> T_C$, i.e. the reconstructed spectrum runs over from the first correlation TOF frame with time off-set $f = 0$ to the second one with $f = 1$.

The situation can be simply summarized as follows. The asynchronous operation of source and statistical chopper makes sure that all chopper time channels $i$ receive counts uniformly integrating the whole source pulse (emulating a continuous source). Thus the measured correlation spectra $M_i(t_S)$ can exactly be converted by eqn. 4 into the direct TOF spectrum $P_j(t_S)$, where $t_S$ stands for a time channel (sub-domain) eventually of a substantial size. This partial direct TOF spectrum will directly reflect the source pulse shape, in particular it will display a sharp cut-off upper limit (corresponding to the starting of the pulse, cf. *b*$^*$*)* diagonal in Figure 4). By the help of eqn. (5) we can will the absolute time scale of partial spectrum $P_j(t_S)$, beyond the folding of the reconstructed spectra into a single time interval $(0, T_C)$ by the correlation algorithm eqn. 4, due to the periodicity of the pseudo-random chopper pulse sequence. Adding together all the partial spectra for all $t_S$ sub-domains covering the whole elapsed time between source pulses, we obtain the full direct TOF spectrum with all effects of the source pulse shape removed averaging (except eventually at the two extreme correlation TOF frames, with $f = 0$ and $f = int(T_S/T_C)$, respectively).

*3.3 Peculiar error propagations*

If we define as a "spectral structure" a single channel in the final reconstructed spectrum, the counting noise, as mentioned above, will be the square root of the integrated spectrum in the $t_s$ source TOF time slice evaluated as an independent block of data. As noted above, this is exactly the counting noise of the sum of all channels in the spectral block, which is by definition the integrated intensity. What happens for spectral structures defined e.g. as the sum of a few neighboring channels between these two extremes is a very important question for understanding the noise propagation behavior in modulation spectroscopy and there is no universal answer. The points (channels) in a reconstructed scattering spectrum $cP_j$ are not statistically independent, since they are computed from a number of independent counts in the modulation spectra, e.g. as various frequency Fourier components of the selected block of the spectrum. Therefore just adding the variances as it is to be done in conventional spectroscopy is not valid; instead one has to follow error propagation through the whole computational process. This can be numerically easily done for each configuration, e.g. for correlation spectroscopy with the *n* = 255 statistical chopper in Figure 2.

The results are shown in Figure 5 for the range of adding the contents of a few channels out of the 255 obtained by taking the TOF spectrum channel equal to the unit time of the modulation, i.e. 1/255 of the period of a full rotation of the chopper. It was further assumed, as most often it is the case that the directly measured modulation spectra $M_i$ contain about the same counts in each channel (an example will be shown below). We find that the variance of the summed up spectra increases much slower than just adding up those for each term in the sum, as in conventional direct spectroscopy. This leads to gains in the data collection rate compared to the conventional case, as shown in the figure. To explain on an example, when we add the contents of two neighboring channels in the final TOF spectrum reconstruction, the noise for the sum remains the same as it was for a single channel (i.e. the data collection rate doubles) and it continues growing more slowly as we increase the number of added channels.

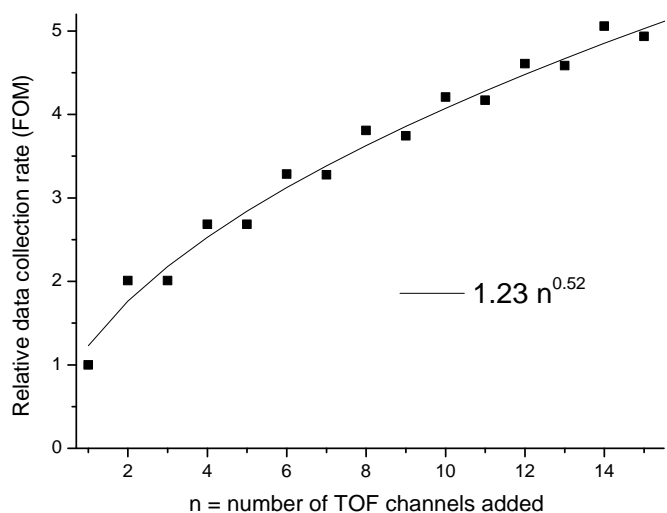

**Figure 5**. The gain in data collection rate (i.e. effective beam intensity) due to noise propagation when adding time channels in statistical chopper based correlation spectroscopy using the *n* = 255 units sequence shown in Figure 2. More details are explained in the text. The line is a guide to the eye and follows the approximate phenomenological formula in the range of added channel numbers shown in the figure.

*3.4 Resolution and intensity*

…..The elementary time unit $t_0$ of the statistical chopper can be considered to be its FWHM TOF time resolution, if the width of the neutron beam window at the chopper is equal to the width of the unit slit. If the width of the beam window is narrower, the effective resolution will only be slightly less than $t_0$, and the beam intensity decreases rapidly. On the other hand, if the width of the beam window is larger than the unit chopper slit, both intensity and FWHM resolution will scale with the beam window width, independently of the width of the unit chopper slit. Slowing down the statistical chopper will not add to the intensity, but will reduce the resolution. Indeed the total intensity transmitted by the statistical chopper is 50 % of the source time average, independently of unit slit width and rotation speed. In fact, from the strict point of view of information contents, broader beam window at maintained high chopper speed does not reduce resolution, since the edges of the resolution line shape remain sharp [17].

….It is important to note that data collection and the correlation data reduction algorithm to be used with reduced resolution by larger beam window (which does not have to be a multiple of the unit slit width) remains the same as in eqn. (4), with the time channel width equal to the unit time of the correlation chopper $t_0$ at its actual speed, independently of the beam widths.

As mentioned in connection with Figure 4, the effective data collection intensity can also be increased by adding neighboring channels in the reconstructed TOF spectrum $P_j$. Here the intensity gain is a computational option and data from a single experimental run can provide both high resolution and high intensity information. This is in full contrast with conventional single pulse spectroscopy, where such options require two separate experimental runs with physically different spectrometer configurations. This de facto adjustment of the spectrometer parameters following the specific experimental needs after completion of the experiment is also new in correlation spectroscopy and it is related to our formulation of correlation spectroscopy in terms of data collection in chopper time unit channels.

*3.5 Correlation frame overlap and source pulse structure*

The rotation period of the statistical chopper implies periodicity of the pseudo-random pulse pattern, which means that the data evaluation cannot distinguish between counts taken with a time difference that is an integer multiple of the chopper revolution time. This phenomenon is a new kind of frame overlap specific for correlation choppers, and it can lead to spurious signals. It can be eliminated by making sure that neutron that can reach the detector at the same time through the correlation chopper cannot have a difference in the time of their crossing of the chopper larger than its full rotation period. At a long pulse source this can be guaranteed if the pulse length of the source, including the pulse tails with significant intensity, is shorter than the rotation period of the statistical chopper. This condition is for example well satisfied by the ESS thermal moderator pulse shape at short wavelengths for a chopper speed of 12000 RPM, i.e. 5 ms per rotation. Figure 6 shows the example of 0.8 Å neutron wavelength ($E_0 = 128$ meV). For higher neutron energies the decay at the tail of the pulse is even faster.

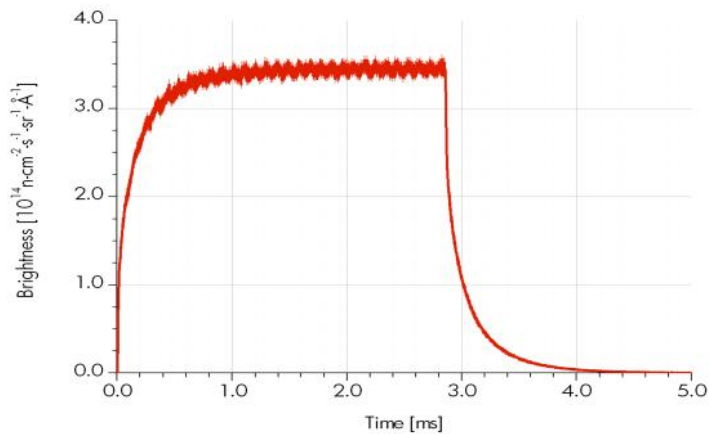

**Figure 6**. Predicted ESS pulse shape at 0.8 Å neutron wavelength.

At longer neutron wavelengths, as we enter more deeply the thermalization regime for $H_2O$ moderators, the pulse decay time constant can approach 0.5 ms and it is not excluded that faster neutrons from the tail of the pulse emitted after 5 ms from the pulse beginning catch up in significant number at the detector the slower ones from the bulk of the pulse. Comparing the total number of neutrons emitted within the first 5 ms long statistical chopper frame to those emitted later, the intensity ratio can be estimated to not to exceed 0.1 %, which might not lead to observable signal above the counting noise. In addition, since the faster neutron signal from the pulse tail will repeat the signal pattern observed 5 ms earlier with neutrons of the same energy from the bulk of the pulse, such an eventual echo delayed by exactly the chopper rotation period can be safely identified and corrected for by software. In any case the chopper rotation period must be significantly longer than the source pulse. Alternatively, for longer neutron wavelengths, depending on instrument geometry, usual frame overlap choppers can also be used to mask the long time tail of the source pulse.

The about 10 KHz systematic noise on the top of the ESS pulse shape shown in Figure 6 is due to the rastering proton beam delivery and need to be considered when assuring asynchronous operation conditions between statistical chopper and source pulse. The rastering modulation can either be in itself synchronous or asynchronous to the source pulse rate. The potential issue can be safely solved by being able to operate the statistical chopper at precisely and freely defined frequencies.

*3.6 Chopper transmission and dynamic range*

It is quite challenging to build disc choppers with higher absorption coefficient than 99.9 % for 500 meV energy neutrons. Such kind of transmission is commonly considered too low for conventional chopper spectrometers and can lead to spurious structures in the spectra. With statistical choppers with 50 % time average transmission such leakage through a closed chopper is of lesser importance than for a single slit chopper that will only be open for typically 1 % of the duration of a long source pulse. The correlation techniques is also systematically little sensitive to background that is not modulated exactly by the pseudo-random pattern of the chopper. Thus as long as the total beam leakage through the statistical chopper blades remains small compared to the 50 % time average "open" transmission (which will determine the counting noise for all time channels in the reconstructed spectra), correlation spectroscopy will work conveniently. This also applies for the fast neutron background following the spallation source pulses.

Thus the operational range of statistical choppers – both from the point of view of sufficient absorption capability and inherently reduced background sensitivity of the correlation technique – can reach as far as 1.5 – 2 eV neutron energy with useful data collection starting about 3 ms after the start of the source pulse. These neutron energies should be made conveniently available by the moderators. The logarithmic (constant $\Delta E/E$) spectral density of $H_2O$ moderators is about constant above 300 meV energy at the level of 10 % of the peak at about 70 meV. Although the solid angle of beam divergence that can be transported by a supermirror neutron guide decreases inversely proportionally to the neutron energy, significant intensity should be deliverable to the sample even in the range of 1 eV neutron energy, in particular in view of the 50 % time average transmission coefficient of the statistical chopper.

## 4. Application for inverted geometry neutron spectroscopy.

The key considerations here are that by two dimensional TOF data collection (i.e. recording events as function of both the time with respect to the source pulse and the asynchronous chopper phase) [15,16], one can apply correlation spectroscopy to slices of spectra collected at times further away from the elastic signal than the source pulse length. For a source like ESS this means that for most of the some 71 ms between pulses the high intensity elastic scattering can be filtered out from the spectra detected in independent $t_s$ sub-domains in order to achieve favorable noise levels, making possible to take good advantage of the very high intensity delivered by 50 % transmission statistical choppers.

We will primarily consider here the use of the correlation method instead of the much lower intensity transmission multichopper system proposed for the variable energy resolution vibrational spectrometer proposal VESPA at ESS, focusing on a broad neutron energy transfer range including high energies up to 0.5 meV, as described in [18].

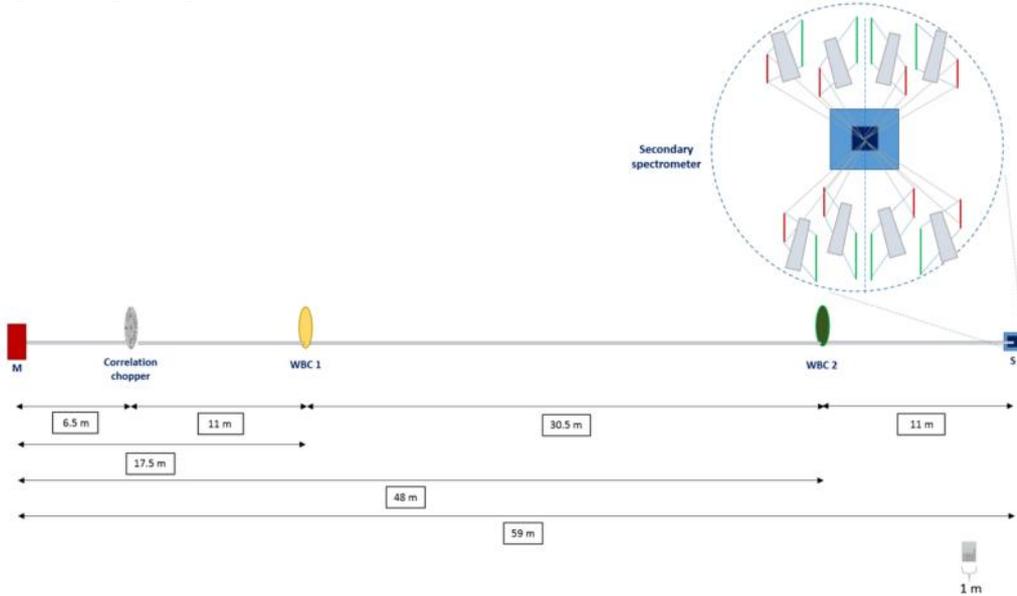

**Figure 7:** The general layout of the VESPA inverted geometry vibrational spectrometer in the alternative proposed here of making use of a statistical chopper for correlation spectroscopy. WBC1 and WBC2 are wavelength band disc choppers with the only task of preventing overlap in the detection of neutrons from subsequent source pulses. The secondary spectrometer [18] contains the crystal analyzers to define the final energy of the neutrons reaching the detectors.

An important point for VESPA and high energy neutron scattering in general is that ESS produces the highest time average hot neutron flux of all spallation sources at all neutron energies, but for good resolution at about 500 meV incoming neutron energy the pulse lengths needs to be cut to some 25 μs, i.e. 99% of the 2.86 ms pulse is discarded. With a correlation chopper 50 % of the time integrated hot neutron flux will be made use of, of course at the price of eventually enhanced counting noise for small signals in the vicinity of large ones, as discussed above.

As shown in Figure 7, the more complex Wavelength Frame Multiplication (WFM) [16] chopper system of VESPA [18] can be replaced by a statistical chopper essentially identical to the one developed and successfully operated at CORELLI (cf. Figure 2 and 3) operating at a reduced speed of 12000 RPM (instead of 18000) and 2-3 simple frame overlap disc choppers. The simulated spectra below are the results of exact analytical modelling, with the only approximations of neglecting the finite manufacturing precision of the statistical chopper and the contribution to the resolution broadening by the secondary spectrometer. The latter was shown to be negligible above 100 meV neutron energy transfer / incoming neutron energy $E_0$ [18]. At the resolution of 1% of $E_0$ planned for VESPA we obtain 90 times more neutrons in the measured spectra, as illustrated by the calculated data in one detector for a mathematical model spectrum, as presented in Figure 8 below.

The "Input" curve is shown with infinite resolution and with the intensity as it would be detected at the WFM chopper version of VESPA. The common time scale for the direct TOF spectrum "input" and the correlation spectra starts at 4 ms after the first chopper pulse in the multichopper configuration of VESPA, to be timed in the middle of the flat top of the ESS source pulse, i.e. about 2 ms after its beginning. The width of the time channels $T_C/n = 19.6$ μs with $T_C = 5$ ms at 200 Hz statistical chopper speed and $n = 255$. The measured correlation spectrum $M_i$ (blue curve) was calculated analytically using eqn. (3). The distribution $t(T_r)$ (which in our geometry can be taken as independent of $r$) is determined by the input model function. The distribution of $t(r)$ was defined by assuming a beam window diaphragm equal to the trapezoidal shape of the single unit length slits of the statistical chopper disc in Figure 2, i.e. of 8.624 mm maximum width at the rim of the chopper disc. As shown in the figure, $M_i$ fluctuates around the time average of the spectrum integrated over the chopper rotation period $T_C$.

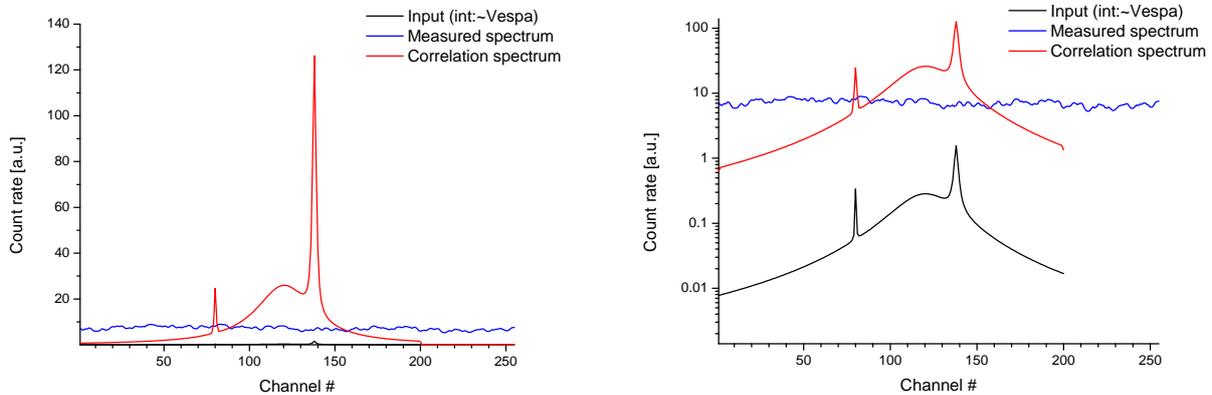

**Figure 8.** Comparison between the observed intensities between VESPA using WFM chopper (black line) and statistical chopper (blue line: measured spectrum, red line: correlation reconstructed direct TOF spectrum). The two plots are identical, one on linear the other on logarithmic scale. More details are given in the text.

The correlation reconstructed TOF spectrum $cP_j$ (red curve) is the raw output of the correlation calculations by applying eqn. (4) to the measured spectrum $M_i$. The constant $c$ comes out to be the product of 127 (in view of the sequence in Figure 2) and the illumination duty factor by the source pulse during the period of the chopper rotation (i.e. 2.86 ms / 5 ms in our case). In the intensity comparison to the input spectrum with VESPA multichopper alternative we have taken into account the pulse intensity at equivalent resolution. The multichopper system as designed in [18] can provide shortest pulses of 49 μs FWHM for the full 3 cm beam width. Tuning the double blind chopper phasing for providing the same 19 μs FWHM resolution reduces both the effective beam width and the peak intensity, which in combination amounts to ¼ of the beam intensity at 30 mm unperturbed beam width. Thus the time average intensity gain by the statistical chopper is a factor of 90 at this highest resolution range $\Delta E/E_0 \sim 0.8$ % at $E_0 = 500$ meV.

The resolution can be reduced for more intensity by extending the width of the beam window and/or by the unique correlation chopper feature of a flexible variety of post-experiment data reduction options for trading resolution vs. intensity by adding neighboring TOF channels, as discussed in connection with Figure 5 above. This latter to some extent has the same impact as the wavelength dependent pulse length by the double blind approach of the multichopper VESPA alternative. But the availability of the highest TOF resolution as data processing option at all incoming neutron energies as defined by the width of the beam window installed is a great advantage, also by the fact that one statistical chopper data collection run can do the job of separate runs using different configurations with conventional direct TOF data collection.

Figure 9 shows the resolution and intensity options obtained by adjusting the beam window in the experiment and adding (grouping) time channels as part of the data analysis after the completed experiment. The wide slit spectra correspond the 3 times wider beam window diaphragm at the chopper compared to the 8.642 mm for the red curve in Figures 8, reproduced in Figure 9 as reference (red curve without symbols). The figure also illustrates the effect of the post-measurement intensity vs resolution trading method: the curves with dots have been obtained by adding 2 neighboring TOF channels and the actual obtained data points with their reduced density are shown as dots in the figure.

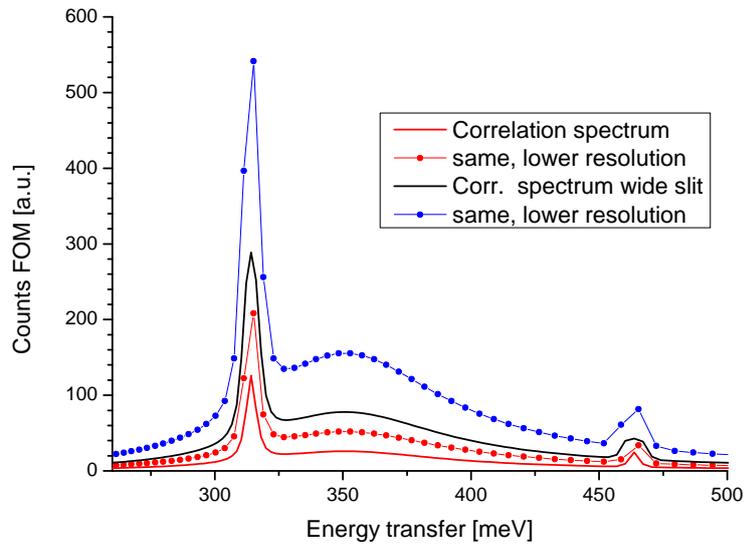

**Figure 9.** Resolution and intensity options by adjusting the beam window in the experiment and adding (grouping) time channels in data analysis after the completed experiment. More details are given in the text.

….All the measured spectrum simulation and its data reduction processing has been done with time channel widths equal to the time unit of the correlation chopper $t_0 = T_C/n$, corresponding to the width of the narrowest slit. This is an important necessity to keep the correlation algorithm exact, as mentioned above, since the $\delta$ function behavior of the autocorrelation function of the chopper modulation applies exactly to the discrete sequence of numbers 0 and 1, cf. Figures 2 and 3 and eqn. 2. It would only approximately apply for a continuous function, which in practical terms would need to be approximated by a fine discrete channel structure for which eqn. (2) would not be valid anymore. The basic correlation formalism in eqns. (2) – (4) is exact, as derived above, if we determine the data points as integrals over time slices of duration strictly equal to the statistical chopper unit time $t_0$. This time interval also exactly corresponds to the FWHM of the shortest pulses produced by the statistical chopper with a beam window width smaller or equal (actually at all radii) to the narrowest chopper slit width.

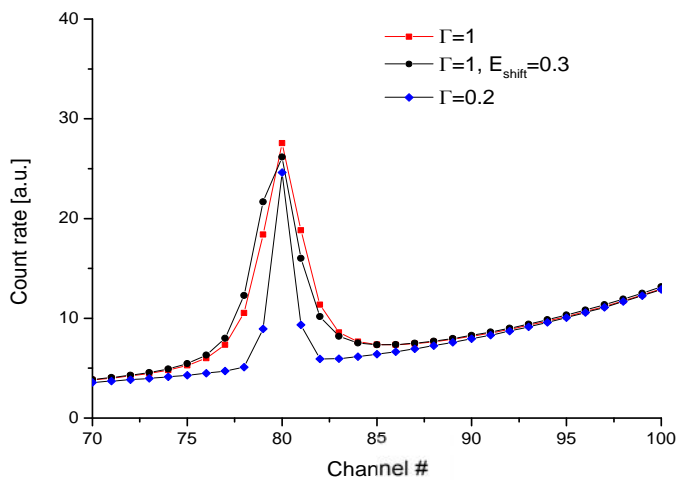

**Figure 10.** Information content of course grain data collection / recording matched to the instrumental resolution as defined by the FWHM $t_0$ of the shortest statistical chopper pulse. The reconstructed correlation spectra in the figure represent Lorentzian lines with half width $\Gamma$ and relative shift in time given in the same units $t_0$ as the time channel width of the spectra shown.

Figure 10 illustrates that such coarse grain data collection does not lead to any loss of information, and well reveals line widths and shifts only amounting to a small fraction of the resolution. Independently of the time channel width, collected spectra reveal little on shapes of lines narrower than the experimental resolution. They remain masked by the broader line shape of the resolution itself. On the other hand, if one is forced to assume a physical line shape due to this masking by the instrumental response, all information on its parameters also can be favorably extracted from a few well defined integrals providing the data points in the coarse grain data collection matched to the instrumental resolution.

Inverted geometry spectrometers are often complemented by detectors with direct view to the sample without analyzer, which deliver the diffraction pattern of the sample measured simultaneously with the inelastic or quasi-elastic spectra. All the considerations above focused on vibrational spectroscopy will also apply to diffraction without any essential change. While diffraction patterns can contain a few reflections with very high intensity compared to the other spectral structures (similarly to the way elastic lines commonly compare to inelastic structure), the effect on data collection counting noise of such large structures will be limited to their vicinity in the TOF spectra with an extension corresponding to the duration of the source pulses. The beam intensity gains by the use of statistical chopper will favorably apply to diffraction data too, in particular by the opportunity of selecting the resolution required in the data processing after the experiments.

## 5. Conclusions

.....We have found that correlation neutron spectroscopy based on the use of pseudo-random statistical beam chopping offers very significant new potentials. These include:
- gains in beam intensity which can approach two orders of magnitude for the study of high intensity structures with dominating contribution to the total scattering intensity of the neutron scattering spectra to be explored;
- gains in beam intensity in the range of an order of magnitude at long pulse neutron sources (or instrumental configurations emulating pulsed source operation on continuous sources) for most parts of neutron scattering spectra by two dimensional time-of-flight data collection for reducing the reach of error propagation over spectral ranges and for extending the band width of correlation data collection to a multiple of the repetition time of the pseudo-random beam modulations;
- enhanced opportunities for variable resolution studies by the unique potential of correlation based data collection for flexibly selecting amongst a variety of balances between resolution and intensity in the data processing after the experiment, i.e. taking advantage of the up to two orders of magnitude enhanced data volume delivered of a single correlation experiment that simultaneously contains high intensity and high resolution information, which can be selectively extracted in the post-experiment data analysis;
- reduced sensitivity to external background which opens up opportunities in high energy neutron spectroscopy with reduced opacity of many equipment components and for the study of weak signals from small samples

We have also identified a formulation of neutron correlation spectroscopy which allows us to exactly reconstruct the direct neutron scattering spectra of physical interest from the enhanced intensity data pseudo-random statistical choppers can provide.

These advantages are particularly relevant for very broad band vibrational neutron scattering spectroscopy including high neutron energies and more generally for inverted geometry spectrometers at the long pulse source ESS.

**Acknowledgements**

Salvatore Magazù and Federica Migliardo gratefully acknowledge financial support from Elettra - Sincrotrone Trieste in the framework of the PIK project "Resolution Elastic Neutron Scattering Time-of-flight Spectrometer Operating in the Repetition Rate Multiplication Mode". Ferenc Mezei gratefully acknowledges the hospitality of Messina University.